\documentclass[%
 reprint,
 amsmath,amssymb,
 aps,
 prb,
 superscriptaddress,
]{revtex4-2}
\usepackage{siunitx}
\usepackage{bbm}

\usepackage{graphicx}
\usepackage[colorlinks=true, allcolors=blue]{hyperref}

\graphicspath{{plots/}}

\date{\today}

\newcommand{\figref}[2][]{%
\ifx\FirstArg\empty
Figure \ref{fig:#2}%
\else
Figure \ref{fig:#2}\subref{#1}%
\fi}

\newcommand{\subref}[1]{\textbf{#1}}
\newcommand{\subf}[1]{\subref{#1}\textbf{,}}

\begin{document}
\title{Low-Energy Electron Microscopy contrast of stacking boundaries: comparing twisted few-layer graphene and strained epitaxial graphene on silicon carbide}
\author{Tobias A. de Jong}
\email{jongt@physics.leidenuniv.nl}
\affiliation{Huygens-Kamerlingh Onnes Laboratorium, Leiden Institute of Physics, Leiden University, Niels Bohrweg 2, P.O. Box 9504, NL-2300 RA Leiden, The Netherlands}
\author{Xingchen Chen}
\affiliation{Huygens-Kamerlingh Onnes Laboratorium, Leiden Institute of Physics, Leiden University, Niels Bohrweg 2, P.O. Box 9504, NL-2300 RA Leiden, The Netherlands}
\author{Johannes Jobst}
\affiliation{Huygens-Kamerlingh Onnes Laboratorium, Leiden Institute of Physics, Leiden University, Niels Bohrweg 2, P.O. Box 9504, NL-2300 RA Leiden, The Netherlands}
\author{Eugene E. Krasovskii}
\affiliation{Departamento de F\'isica de Materiales, Universidad del Pais Vasco UPV/EHU, 20080 San Sebasti\'an/Donostia, Spain}
\affiliation{IKERBASQUE, Basque Foundation for Science, E-48013 Bilbao, Spain}
\affiliation{Donostia International Physics Center (DIPC), E-20018 San Sebasti\'an, Spain}
\author{Ruud M. Tromp}
\affiliation{IBM T.J.Watson Research Center, 1101 Kitchawan Road, P.O.\ Box 218, Yorktown Heights, New York 10598, USA}
\affiliation{Huygens-Kamerlingh Onnes Laboratorium, Leiden Institute of Physics, Leiden University, Niels Bohrweg 2, P.O. Box 9504, NL-2300 RA Leiden, The Netherlands}
\author{Sense Jan van der Molen}
\email{molen@physics.leidenuniv.nl}
\affiliation{Huygens-Kamerlingh Onnes Laboratorium, Leiden Institute of Physics, Leiden University, Niels Bohrweg 2, P.O. Box 9504, NL-2300 RA Leiden, The Netherlands}

\begin{abstract}
Stacking domain boundaries occur in Van der Waals heterostacks whenever there is a twist angle or lattice mismatch between subsequent layers. Not only can these domain boundaries host topological edge states, imaging them has been instrumental to determine local variations in twisted bilayer graphene.

Here, we analyse the mechanisms causing stacking domain boundary contrast in Bright Field Low-Energy Electron Microscopy (BF-LEEM) for both graphene on SiC, where domain boundaries are caused by strain and for twisted few layer graphene. We show that when domain boundaries are between the top two graphene layers, BF-LEEM contrast is observed due to amplitude contrast and corresponds well to calculations of the contrast based purely on the local stacking in the domain boundary. 
Conversely, for deeper-lying domain boundaries, amplitude contrast only provides a weak distinction between the inequivalent stackings in the domains themselves. However, for small domains phase contrast, where electrons from different parts of the unit cell interfere causes a very strong contrast.
We derive a general rule-of-thumb of expected BF-LEEM contrast for domain boundaries in Van der Waals materials.
\end{abstract}

\maketitle
\section{Introduction}

Multiple layers of graphene can exist in several stable stacking configurations. Both in twisted heterostacks of 2 or more layers of graphene and in systems on a substrate where different orientations can nucleate and/or relative strain can exist, domains of different stacking configurations co-exist. 

The existence of different domains and the spatial variability of such domains turn out to be essential to explain the (variation of) electronic properties in heterostructures of Van der Waals materials in general.
In particular, when the lattice mismatch is small, any small variation in the atomic lattice is strongly magnified in the superlattice, which in turn determines the electronic properties. 
For example, in twisted bilayer graphene (TBG) the variations in strain change the superconducting properties~\cite{mesple_heterostrain_2021,de_jong_imaging_2021,halbertal_moire_2021,benschop_measuring_2021,kazmierczak_strain_2021}. For larger domains, the domain boundaries themselves host topological boundary states~\cite{martin_topological_2008,huang_topologically_2018,verbakel_valley-protected_2021,yin_direct_2016}.
Imaging the precise stacking and the domains is therefore not only a way to accurately measure the local atomic lattice mismatch and to image topological atomic defects~\cite{de_jong_imaging_2021,ravnik_strain-induced_2019}, but crucial to understand electronic properties heterostructures of multiple graphene layers, in 1-on-1 TBG, but in particular also in thicker twisted samples which are gaining in relevance, such as 2-on-2 TBG and beyond.

In this work, we use Bright Field Low Energy Electron Microscopy (BF-LEEM) to characterize the contrast of domain boundaries in both twisted graphene systems and in the strained graphene on SiC and subsequently compare them~\cite{Tromp2010,tromp2013new,dejong2018intrinsic,de_jong_imaging_2021}.
In previous work, we have shown that Dark Field LEEM can be used to image stacking domains in bilayer and trilayer graphene on SiC~\cite{dejong2018intrinsic}. 
Tilted DF-LEEM was used, as the rotational equivalency between AB and AC stacking means no contrast can be expected in Bright Field LEEM.
However, the domain boundaries themselves can be imaged in BF-LEEM, as was previously demonstrated for the case of twisted bilayer graphene in Ref.~\cite{de_jong_imaging_2021}.

Here, we focus on the precise contrast mechanisms enabling this. To do so, the intensity of the domain boundaries needs to be separated from the domains themselves, which is non-trivial because the domain boundaries are about 10\,nm wide at most. Improving on a PCA-based method used for this goal in Ref.~\cite{de_jong_quantitative_2020}, we here average over multiple unit cells to increase the resolution and signal-to-noise ratio and extract the contrast information as a function of $E_0$ from an averaged unit cell. 
First, we discuss the material systems and precise type of domain boundaries occurring in them, then the averaging method and the results.

\subsection{Graphene on silicon carbide}

Graphene on silicon carbide (SiC(0001)) is grown by thermal decomposition. 
The main advantage of this growth method is that the growth is epitaxial, and results in a single orientation of graphene.
However, the lattice constant of hexagonal carbon does not match the lattice constant of SiC. Thus a higher-order commensurate reconstruction, i.e. a moir\'e pattern is formed, denoted by $(6\sqrt3 \times 6\sqrt3)R30^\circ$~\cite{riedl_structural_2007,kim_origin_2008}.

The first layer of hexagonal carbon is covalently bonded to the SiC surface. This means this so-called buffer layer is insulating due to the lack of pure sp$^2$ hybridization and that it is strained somewhat (compared to pure graphene), to perfectly adhere to the higher-order commensurate reconstruction.
All subsequent carbon layers are true graphene layers and thus only bonded to the lower layers by Van der Waals forces. 
Aside from the implications for the conduction, this also implies that the graphene layer on top of the buffer layer has much lower interlayer interaction energies than the buffer layer with respect to the SiC substrate.
Therefore, the graphene can assume its own lattice constant, with any residual lattice constant mismatch between the graphene and the $(6\sqrt3 \times 6\sqrt3)R30^\circ$ reconstruction of the buffer layer resolved, especially at the high growth temperatures.

\begin{figure}
\includegraphics[width=\columnwidth]{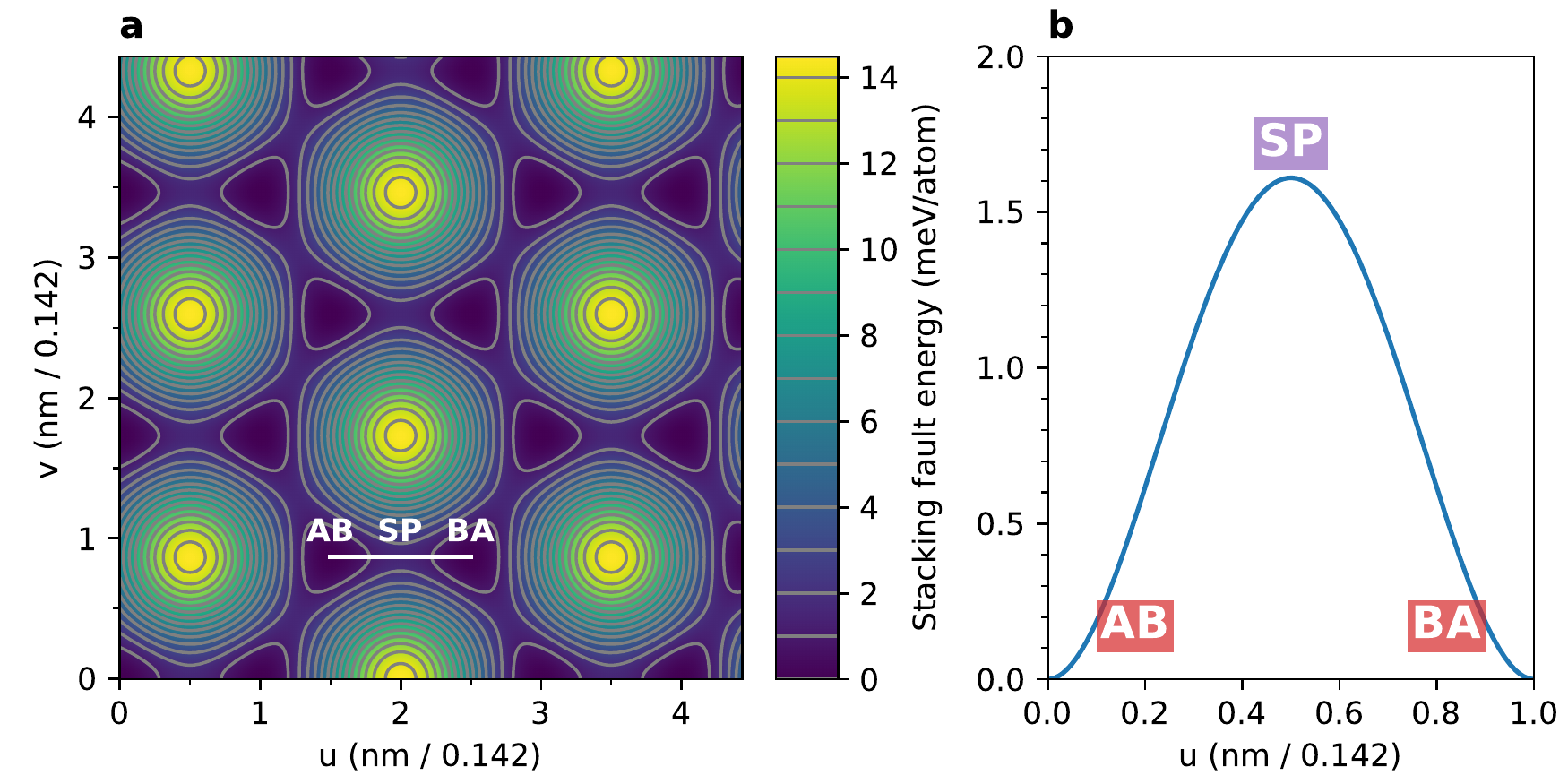}
\includegraphics[width=\columnwidth]{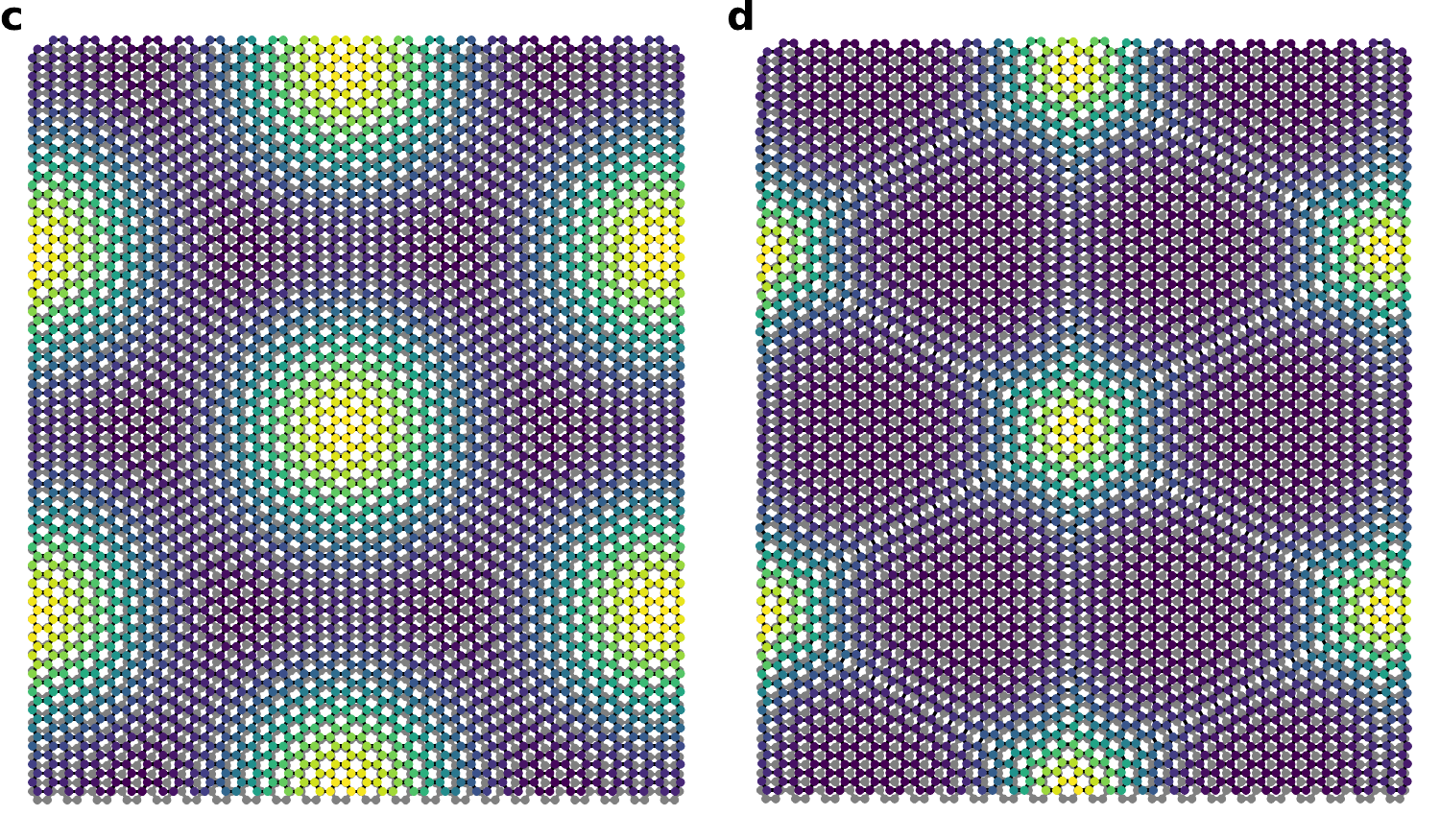}
\caption{\subf{a} Approximate interlayer stacking energy (in meV/atom) for bilayer graphene as a function of relative displacement in units of the graphene bond length $l_0=0.142\,\text{nm}$, as given in Ref.~\cite{lebedeva_two_2020}. (c.f. calculated energy in Ref.~\cite{popov2011commensurate}). Note that the stacking energy for AA stacking relative to Bernal stacking (AB/BA) is around 9 times higher than the maximum occurring in a domain wall, which is labeled SP (for saddle point). 
\subf{b} Least energy cut through the energy landscape as indicated in \subref{a}, from AB to BA stacking across the saddle point.
\subf{c} Schematic of two unrelaxed hexagonal lattices with slightly different lattice constant, where the color of each atom indicates the stacking fault energy (as in \subref{a}). 
\subf{d} As Bernal stacking (AB/BA) is energetically favorable compared to other stackings, the bilayer will relax to form triangular Bernal stacked domains with all strain concentrated in the boundaries, resulting in much lower total stacking fault energy.}
\label{fig:stacking_domains}
\end{figure}

As the buffer layer is similar to a graphene layer, the interlayer stacking energy landscape should be similar to that of bilayer graphene, which is shown in \figref[a]{stacking_domains}~\cite{lebedeva_two_2020}. Here, the Bernal stackings (AB/BA) are the energy minima. 
When one of the layers is shifted to form AA stacking, this corresponds to a maximum. 
For a small residual lattice mismatch, schematically shown in \figref[c]{stacking_domains}, the relative stacking and therefore the local interlayer stacking energy varies continuously as a function of position.
When relaxing this structure, the interlayer stacking energy will be minimized at the cost of some stretching of the layer. Now, triangular domains form, where in each boundary the strain is concentrated (\figref[c]{stacking_domains}), and the stacking varies smoothly, going from one Bernal minimum to the other via the saddle point (SP) in the energy landscape (\figref[b]{stacking_domains}).

Indeed, we have shown in earlier work that such stacking domains form~\cite{carr_relaxation_2018,annevelink_topologically_2020}, with an influence on the (de-)intercalation process~\cite{dejong2018intrinsic}.

\subsection{Twisted few-layer graphene}
In twisted few-layer graphene made by mechanical exfoliation and re-assembly, the lattice mismatch is not due to an intrinsic mismatch of the lattice constant of the graphene with respect to that of the substrate, but by artificially rotating the top layers by a twist angle $\theta$ with respect to the bottom layers.

Here, a continuous transition from the commensurate case at $\theta=0$ to the incommensurate case for twist angles larger than a critical angle occurs. This critical angle depends on the precise number of layers. Here, precise estimates of the critical angle vary, with estimates for the 1-on-1 layer case between about $1$ and $2$ degree~\cite{Yoo2019,carr_relaxation_2018}. Notably, this critical angle and the first magic angle for bilayer graphene are very close. What is more, for additional layers, both angles increase. Below the critical angle, locally commensurate stacking domains form, with all strain concentrated in domain boundaries~\cite{carr_relaxation_2018,annevelink_topologically_2020}. 
However, these domain boundaries are qualitatively different for the twisted case compared to the biaxially strained case: while in the strained case the lattice mismatch or displacement compensated by the domain boundary is perpendicular to the so-called \textbf{tensile} domain boundary, in the twisted case this is parallel to the so-called \textbf{shear} domain boundary.

In the more general case of mixed twist and (uniaxial) strain, mixes between these two types also occur.
Applying the two-chain Frenkel-Kontorova model to bilayer graphene, Lebedeva and Popov found that the shear domain boundary has a slightly lower total energy cost per unit length than the tensile boundary~\cite{lebedeva_two_2020}. They also calculated a width of 13.4\,nm for the tensile domain wall and 8.6\,nm for the shear domain wall. These values match experimental values of 11\,nm and 6--7\,nm measured using TEM~\cite{alden2013strain,lin_acab_2013} to within the expected accuracy of their model.

Both in the twisted case and in the biaxially strained case, domain boundaries that occur at different azimuthal angles have to cross. In bilayer graphene, such a domain boundary crossing corresponds to AA-stacking, and is therefore called an AA-node.

Notably, in the twisted case, such domain boundary stackings are in some sense topologically protected: short of destroying the lattice by adding or removing atoms, they can only be destroyed by moving them all the way to the edge of the system. As they therefore exhibit particle-like properties, they are sometimes called \textbf{twistons}~\cite{turkel_twistons_2021}.
Similar properties hold in the strained case, and therefore we mint the term \textbf{strainons} for AA-nodes in graphene on SiC.

\section{Stacking contrast of bilayers in LEEM}

\begin{figure}[!ht]
\includegraphics[width=\columnwidth]{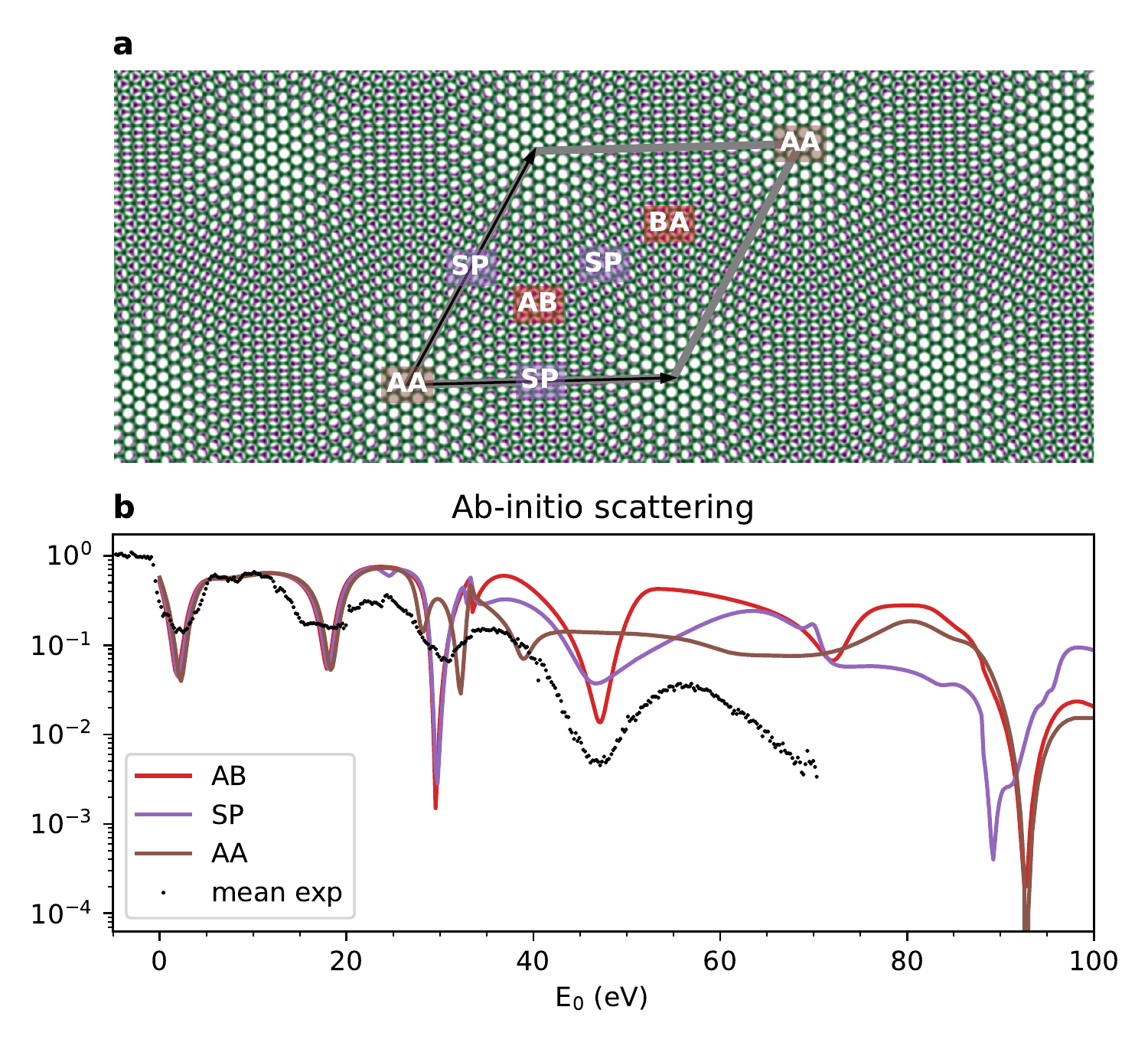}
\caption{
\subf{a} Schematic of a TBG unit cell schematic with different local relative stackings labeled at the positions in the unit cell where they occur. 
\subf{b} Calculated electron reflectivity with the ab-initio Bloch-wave-based scattering method described in Ref.~\cite{krasovskii_ab_2021} for different bilayer graphene stackings. In black dots, the mean reflectivity of $\theta\approx 0.18^\circ$ TBG on hBN is overlayed for comparison.
}\label{fig:stackingcalculations}
\end{figure}

In Dark Field LEEM (DF-LEEM), the rotational equivalence between the two possible Bernal stackings, AB and BA, is broken, causing contrast between the domains themselves~\cite{dejong2018intrinsic,schadlich_stacking_2021}. 
In BF-LEEM, both Bernal stackings are fully equivalent by rotation and no contrast between them can be expected, but the domain boundaries themselves do cause contrast.
To understand the domain boundary contrast observed with LEEM we would like to compare measurements to theoretical calculations. 
Unfortunately, the super cells, both of twisted bilayer graphene at angles near the magic angle and of any reasonable lattice mismatch caused by strain, contain too many atoms to be amenable to reflectivity calculations using conventional methods. 
A simplifying assumption to tackle the problem would be that for large enough unit cells, the main contrast mechanism is due to stacking contrast, e.g. the different local stackings in the super cell having slightly different electron reflectivities as a function of landing energy, causing visible contrast to image the super cells.
Here any lateral interaction between the different areas in the moir\'e unit cell is ignored, which is equivalent to assuming pure amplitude contrast and no phase contrast~\cite{schramm_contrast_2012}.

To test this assumption, we compare experimentally observed contrast to ab-initio calculations from different sources: an ab-initio Bloch-wave-based scattering method~\cite{de_jong_imaging_2021,de_jong_data_2021} and traditional tensorLEED calculations as reported in Ref.~\cite{hibino_stacking_2009}.
Computed reflectivity curves for the Bloch-wave-based method are shown in \figref{stackingcalculations}, together with an indication of where the different stackings occur in the unit cell of TBG.

Both these calculations and the tensorLEED calculations in Ref.~\cite{hibino_stacking_2009} predict very little contrast between different stackings at landing energies lower than the appearance of the first order diffraction spots, i.e. $E_0 \lesssim 30$\,eV. The contrast increases for higher $E_0$. 
However, two things should be noted here. 
First, The ab-initio scattering method is much more accurate at low energies, as the so-called muffin tin approximation used in tensorLEED severely limits its accuracy in this energy regime. 
Remarkably, at these low energies, the difference between the different stackings in the ab-initio scattering calculations seems to be limited to a small shift along energy, i.e. a slight work function difference.
The second thing to note is that although high contrast is predicted for higher energies, in experimental practice, the measured contrast for higher energies is decreased by both inelastic losses, causing broadening of the measured spectra, and decreasing intensity, causing decreased signal-to-noise ratios. 
This means that a priori, it is not clear from these calculations what would be the optimal energy to measure such stacking contrast.

\subsection{Unit cell averaging}

To further complicate comparison to experiment, the width of a single domain boundary is too small to accurately sample at a single position, making comparison to the calculated reflectivity of different stackings for different regions of interest impractical.\footnote{In fact, the domain boundaries might be too thin to observe at all in non-aberration corrected LEEM, as attempts using microscopes without aberration correction have so far been unsuccessful.}

\begin{figure}[!ht]
\includegraphics[width=\columnwidth]{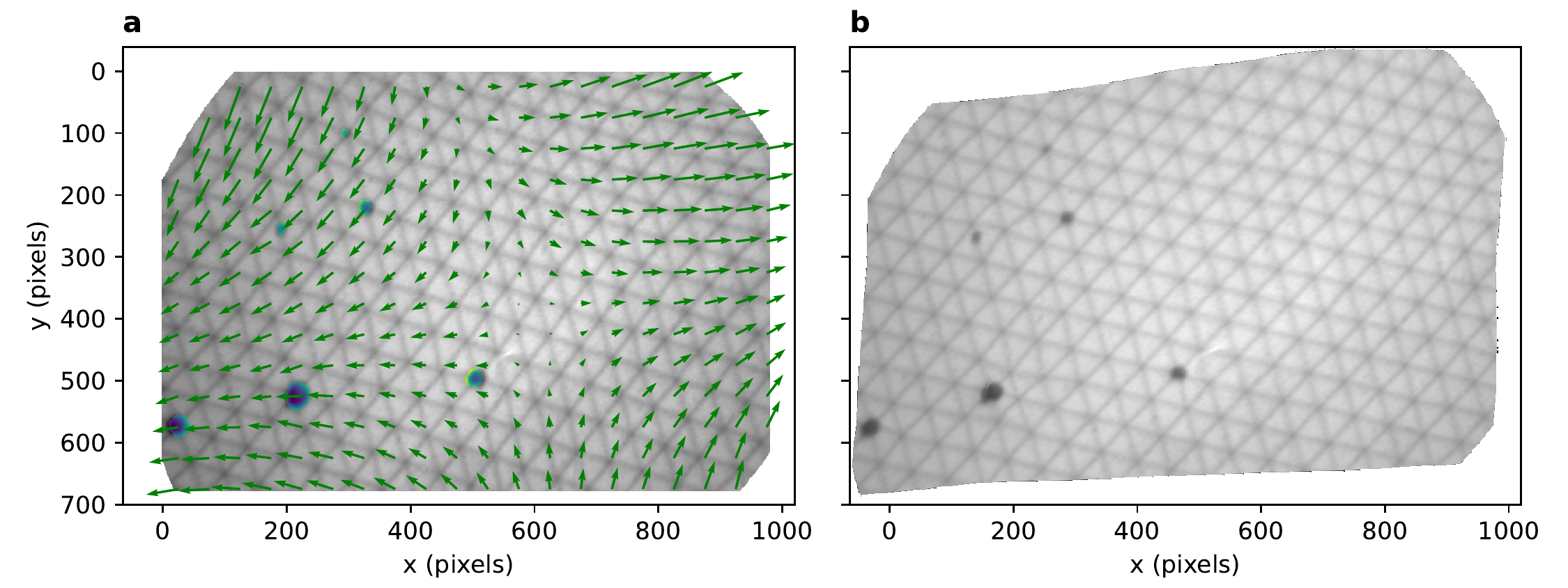}
\includegraphics[width=\columnwidth]{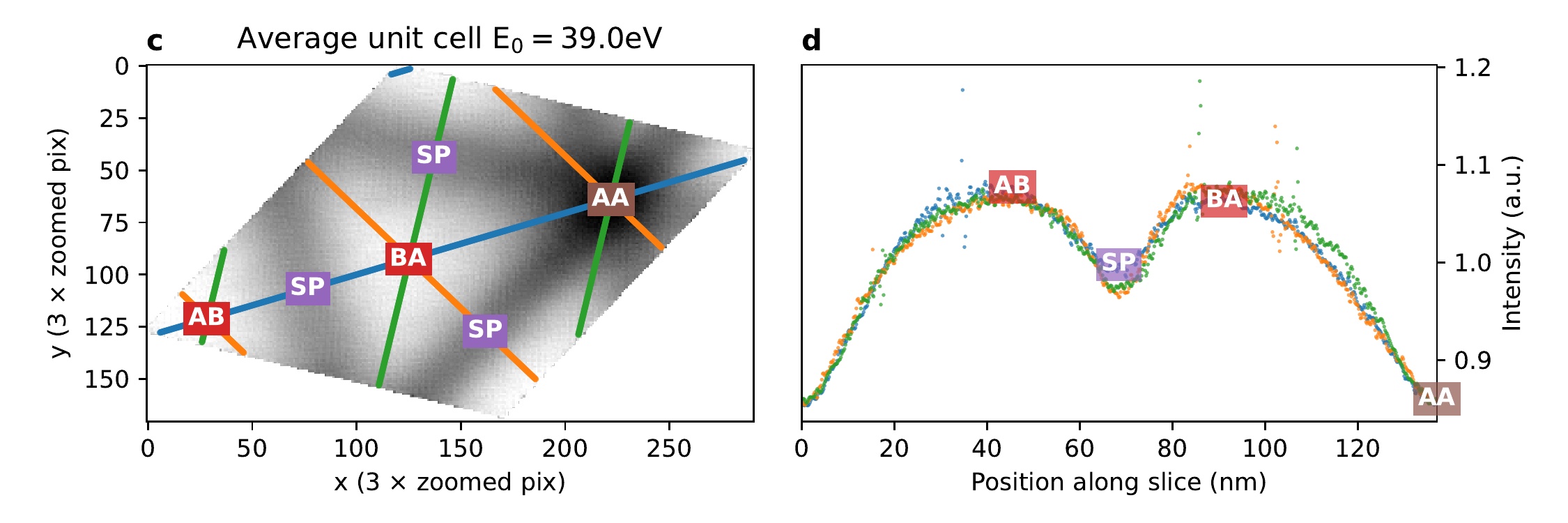}
\caption{\textbf{Unit cell averaging.}
\subf{a} From a displacement field $\vec u (\vec r)$ calculated using geometric phase analysis (green arrows), we can compute a corrected regular lattice as shown in 
\subref{b}. This regular lattice can be averaged by projecting into a single unit cell \subref{c} by subtracting integer multiples of the lattice vectors.
\subf{d} To visualize the unit cell as a function of energy, equivalent cuts in different directions along the unit cell can be made. Colors match the indicated slices in \subref{c} and \figref{TBGcuts}. Spikes in the intensity are due to incorrect handling of the edges of the unit cell and are filtered out in the results.
The image used for illustration here is the $\theta \approx 0.18^\circ$ TBG sample also used in \figref{TBGcuts}.
}
\label{fig:unitcellaveragingexpl}
\end{figure}

Therefore, to optimally compare experiment and theory, we will try to average data over multiple unit cells of the moir\'e lattice. However, in general, strain and twist angle variation will cause deformation of the unit cell, which means we can not just project back into the unit cell by shifting pixels over integer multiples of the unit vectors. 
Instead, as illustrated in \figref{unitcellaveragingexpl}, we should first correct the deformation due to strain and twist angle, which we can do by calculating the displacement field $\vec u (\vec r)$ (green arrows in \figref[a]{unitcellaveragingexpl}) using geometric phase analysis (GPA)~\cite{de_jong_imaging_2021,benschop_measuring_2021}, such that $\vec{r'} = \vec r +\vec u (\vec r)$ with $\vec{r'}$ the corresponding position in the \textit{undistorted} lattice.
This can then be used to perform a Lawler--Fujita type distortion correction~\cite{de_jong_imaging_2021,benschop_measuring_2021,slezak_imaging_2008,Lawler2010}, where an undistorted image is sampled from positions $\vec{r'} + \vec u^{-1}(\vec{r'})$ by interpolation, where $\vec u^{-1}$ is determined by approximation or by numerical inversion. 
Of the resulting image, shown in \figref[b]{unitcellaveragingexpl}, it is now possible to project all cells into a single unit cell (\figref[c]{unitcellaveragingexpl}) by integer multiples of the unit vectors:
\[\vec r_p = \left(A^{-1} \vec{r}\right)\mod 1 \]
Here, $A$ is the matrix with the lattice vectors as columns, such that $A^{-1}$ converts to coordinates in terms of the lattice vectors.

However, this two-step process would cause interpolation errors twice and is unsuited for upscaling of the unit cell to recover more detail. Fortunately, once $\vec u (\vec r)$ is known, we can directly compute the precise (i.e. sub-pixel coordinates) position inside the unit cell for each pixel in the original image: 

\[\vec r_p = \left(A^{-1} (\vec r +\vec u (\vec r))\right)\mod 1\]

Therefore we can directly combine all pixels of the original image (\figref[a]{unitcellaveragingexpl}) into an average unit cell (\subref{c}), scaling up and using a `drizzle'-like approach~\cite{fruchter2002drizzle,quist_superresolution_2020} to minimize the smoothing caused by the recombination and we may even hope to recover some additional detail not apparent from the original images.\footnote{The amount of detail within the unit cell that can be recovered in this way depends on the ratio between the pixel pitch and the width of the contrast transfer function (CTF) of the instrument. Therefore this technique might be applied with much more result to experiments where this ratio is large, such as large field-of-view STM, STEM, or AFM measurements.}

The process described above allows us to compute a single average unit cell from an image with distortion, provided that the moir\'e contrast and signal-to-noise ratio are high enough. By doing this for all images in a spectroscopic LEEM dataset, we can obtain the average unit cell reflectivity as a function of $E_0$.
However, the contrast of the moir\'e will be essentially zero for some energies, causing the extraction of the distortion field to fail. 
We also need to exclude areas with significant adsorbates. Furthermore, the area used for averaging should be limited to an area with approximately constant distortion, as the contrast may depend on the distortion. For example, domain boundaries have an approximately constant width, independent of unit cell size and distortion, which is thus distorted when projecting back different size unit cells to a single unit cell.
Accommodating these complications, the unit cell averaging process we use is as follows:

\begin{enumerate}
\setcounter{enumi}{-1}
\item Properly correct the dataset for detector artefacts and drift.
\item Compute $\vec u (\vec r)$ with respect to an isotropic lattice for a value of $E_0$ where the contrast of the moir\'e is high enough. Preferably use an image consisting of the average over a few images around that energy to minimize noise.
\item Determine the high symmetry point (in practice, we select the AA site by finding the minimum or maximum in the unit cell) from the same image. This is used to take one-dimensional slices of the data later on.
\item Mask out any adsorbates and otherwise unwanted areas to be explicitly ignored in the actual unit cell averaging.
\item Use the same distortion field $\vec u (\vec r)$ to compute an average unit cell for all landing energies.
\item Take appropriate slices through the average unit cells that enumerate the theoretically computed stackings (blue, orange, and green lines in \figref[c]{unitcellaveragingexpl}).
\item To cancel out disagreements between models and experimental data in the global intensity, divide these cuts by some reference stacking, in this case Bernal stacking. In the following, if this is not feasible due to remaining detector drift, we divide by the average spectrum instead. Finally, for comparison, we take the natural logarithm of the result.
\end{enumerate}

The core unit cell averaging algorithm is written in Python and made available as part of \texttt{pyGPA}~\cite{pyGPA}, and the Python code used to generate the figures in this work is available at Ref.~\cite{graphene-stacking-domains-code}.

\subsection{Twisted Bilayer Graphene results}

\begin{figure*}[!ht]
\includegraphics[width=\textwidth]{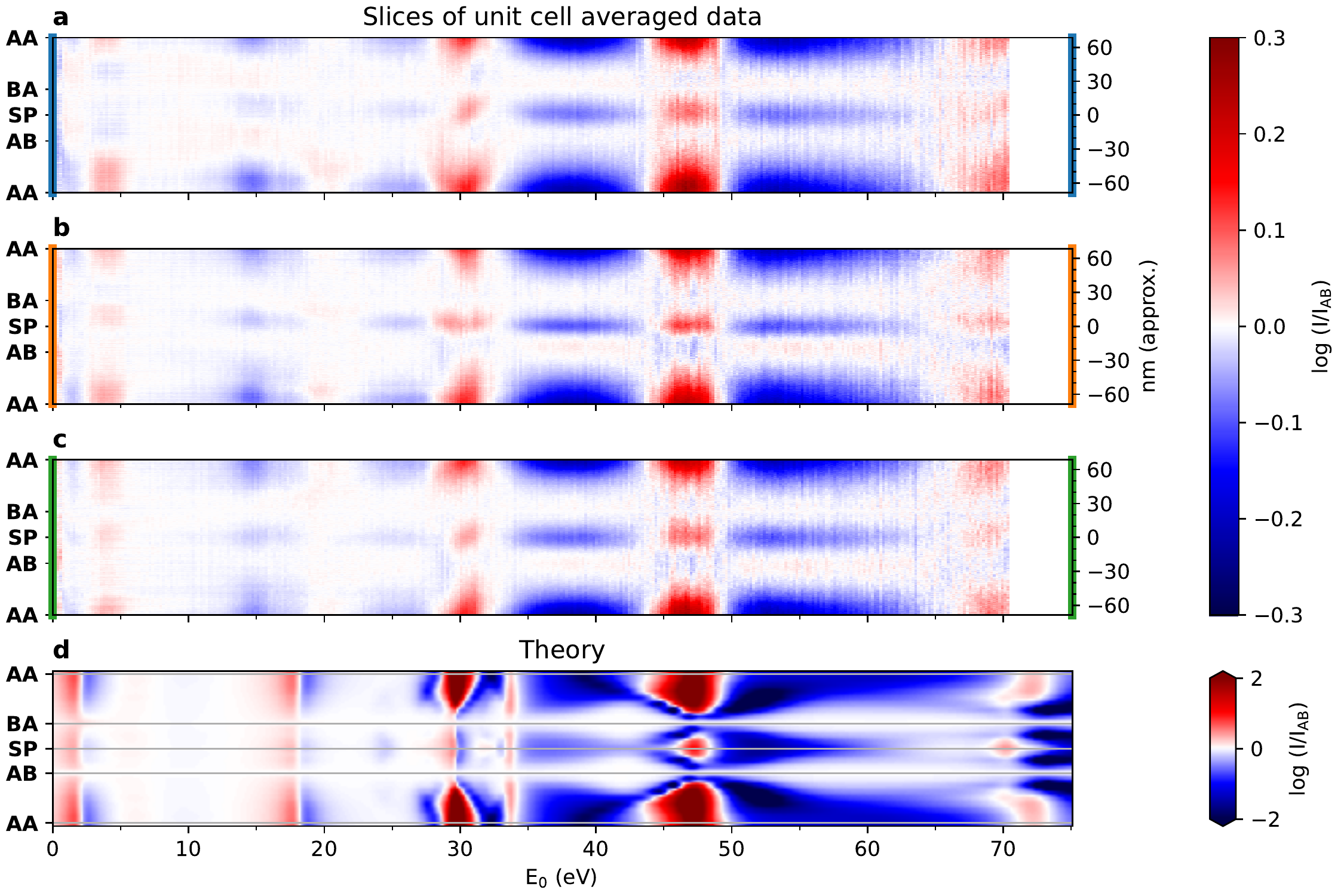}
\caption{\subf{a-c} Natural logarithm of the intensity of cuts through the averaged unit cells normalized with respect to the Bernal reflectivity in the three equivalent directions indicated with the same colors in \figref{unitcellaveragingexpl}c (corresponding to $\theta \approx 0.18^{\circ}$, detector resolution was 1.36\,nm/pixel.).
\subf{d} Calculations of shifted equivalent stackings using the ab-initio theory, smoothed with a Gaussian with $\sigma= 0.2$\,eV to account for experimental smoothing.
}\label{fig:TBGcuts}
\end{figure*}

The unit cell averaging procedure introduced in the previous section is applied to a dataset of twisted bilayer graphene (TBG), with a twist angle of $\theta \approx 0.18^{\circ}$ and a detector resolution in the original dataset of 1.36\,nm/pixel (See \figref{unitcellaveragingexpl}).
The results are compared to the ab-initio theory in \figref{TBGcuts}. Although the experimental contrast is much lower, a remarkably good correspondence of the qualitative features is achieved above $20$\,eV. This includes the contrast inversions, where domain boundaries and the AA site change from brighter than the Bernal (AB or BA) stacking (red) to darker (blue) and vice versa as a function of energy.

Therefore, we conclude that at low twist angles, the moir\'e contrast is mainly caused by the different electron reflectivity of different local stackings and no significant phase contrast plays a role.

However, limitations of this approach in its current form are also immediately visible. Around contrast inversions, most prominently around 30\,eV, it is clear from the asymmetric and different shapes in the three slices that the drift correction was not perfect, even relative to the large unit cell of this low twist angle.
Note that the contrast inversions take place around the minima of the original spectra, where low intensity and energy spread of the electron source cause the most significant artefacts.

Notably, for lower energies, where ab-initio scattering mostly predicts a slight shift along $E_0$, experiment seems to indicate the inverse contrast, i.e. a shift in the opposite direction.

\begin{figure}[!ht]
\includegraphics[width=\columnwidth]{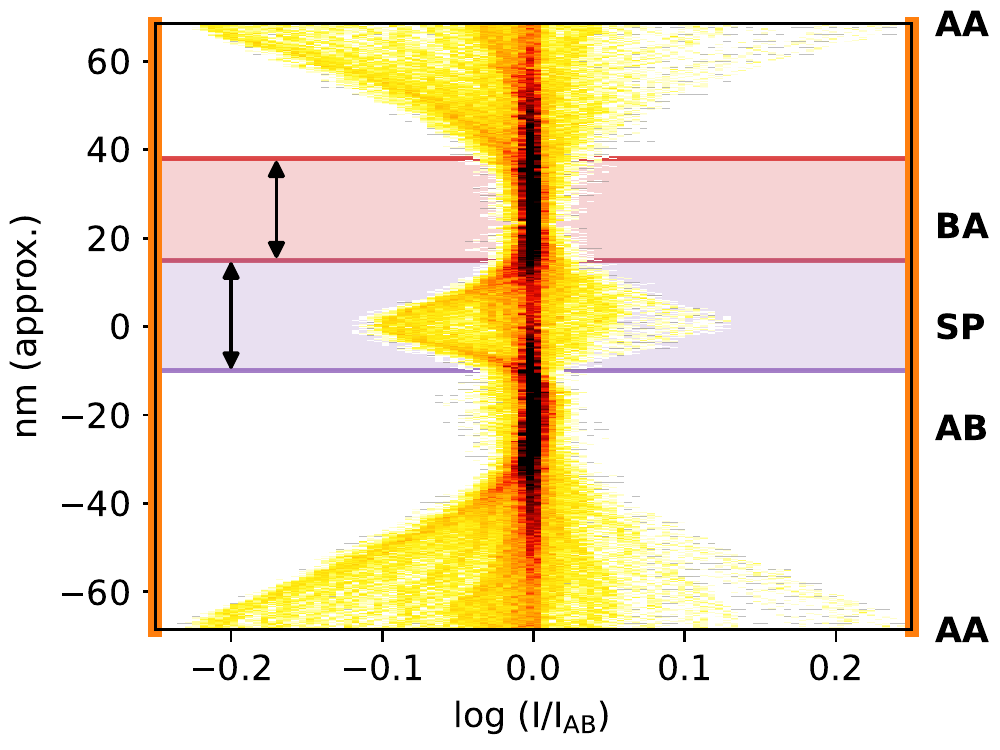}
\caption{Histogram of the relative intensity (with respect to AB stacking) in \figref[b]{TBGcuts}, projected along $E_0$. Indicated in purple is the extracted domain boundary width, and in red the Bernal stacked area.
}\label{fig:TBGprojs}
\end{figure}

Real space dimensions can also be extracted from these slices.
The width around the indicated Bernal stacking in \figref{TBGcuts} with approximately the same intensity is significantly larger than in the theoretical curves. This reaffirms that relaxation to Bernal stacking takes place, forming locally commensurate domains~\cite{Yoo2019,carr_relaxation_2018} (which was also clear from the original data, such as in \figref[a,b]{unitcellaveragingexpl}).

This broadening can be observed more clearly from the histogram of log-contrast values projected along $E_0$, as shown in \figref{TBGprojs}.

The width of the domain boundary is extracted from this, by measuring the length along the cut between AB and BA which has (significant) deviation from the Bernal stacking intensity for the full range of $E_0$, as indicated in in \figref{TBGprojs}.
The observed width of about 25\,nm is still much higher than the expected 7\,nm, possibly by smearing during unit cell averaging, both intrinsic (thermal) broadening and electron optical broadening, and from imperfections of the extracted $\vec u (\vec r)$.

\subsection{Comparison to strain domain boundaries in graphene on SiC}

Next, we compare the results on TBG from the previous section to the domain boundaries as observed in epitaxial graphene on silicon carbide. In the latter case, intrinsic stacking domains occur due to the lattice mismatch between the buffer layer and the graphene layers~\cite{dejong2018intrinsic}.
This means that in this system stacking contrast should occur due to tensile domain boundaries. This should hold both for hydrogen intercalated graphene on SiC, so-called quasi-freestanding bilayer graphene (QFBLG), and for epitaxial monolayer-on-buffer layer in the non-intercalated or de-intercalated material (EMLG). Indeed, domain boundaries in both systems cause contrast in BF-LEEM. 
However, due to intrinsic disorder in this system~\cite{de_jong_2022_morphology}, no areas were imaged that are homogeneous enough to apply GPA to enable the same unit cell average analysis as applied in the previous section. 

\begin{figure}[!hb]
\includegraphics[width=0.9\columnwidth]{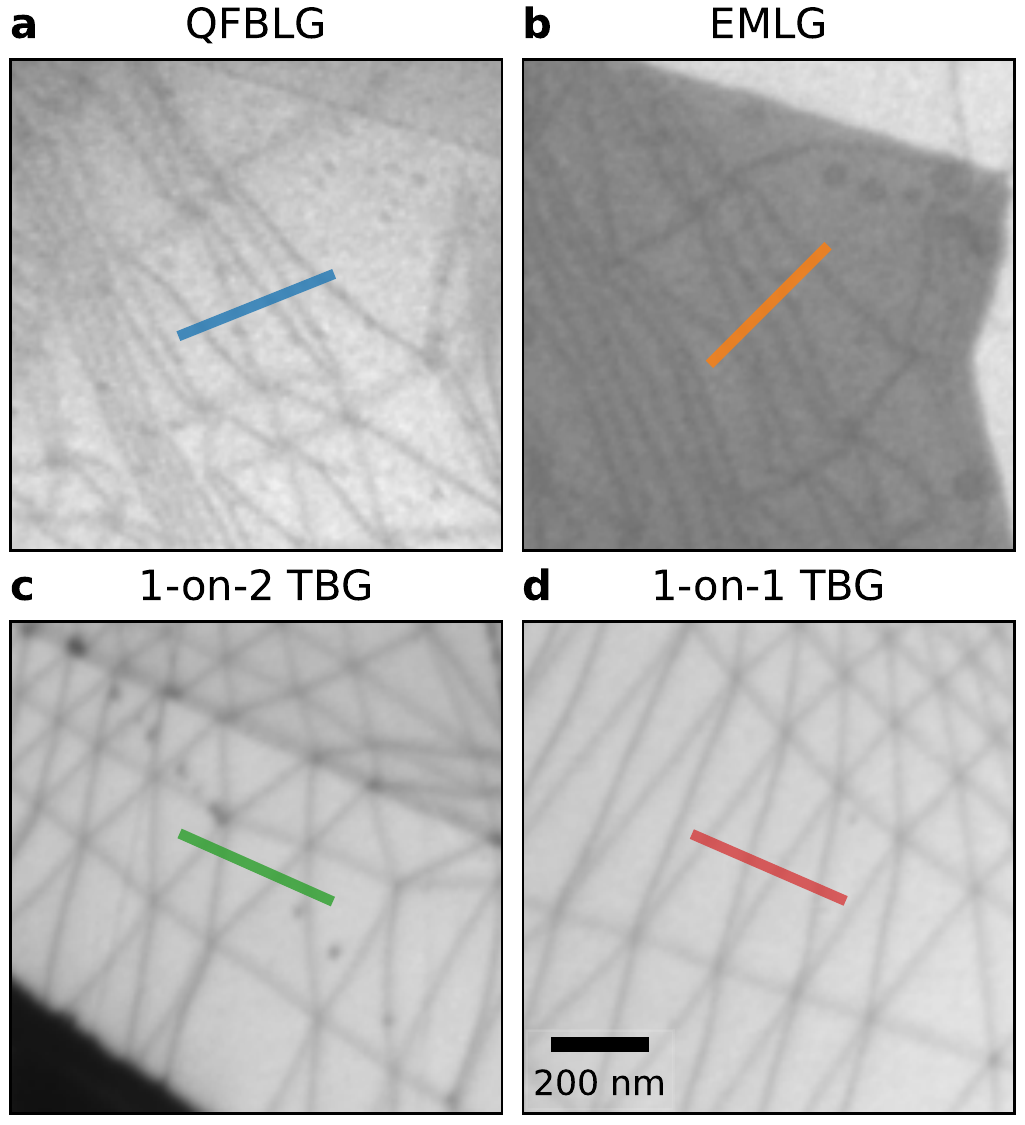}
\includegraphics[width=\columnwidth]{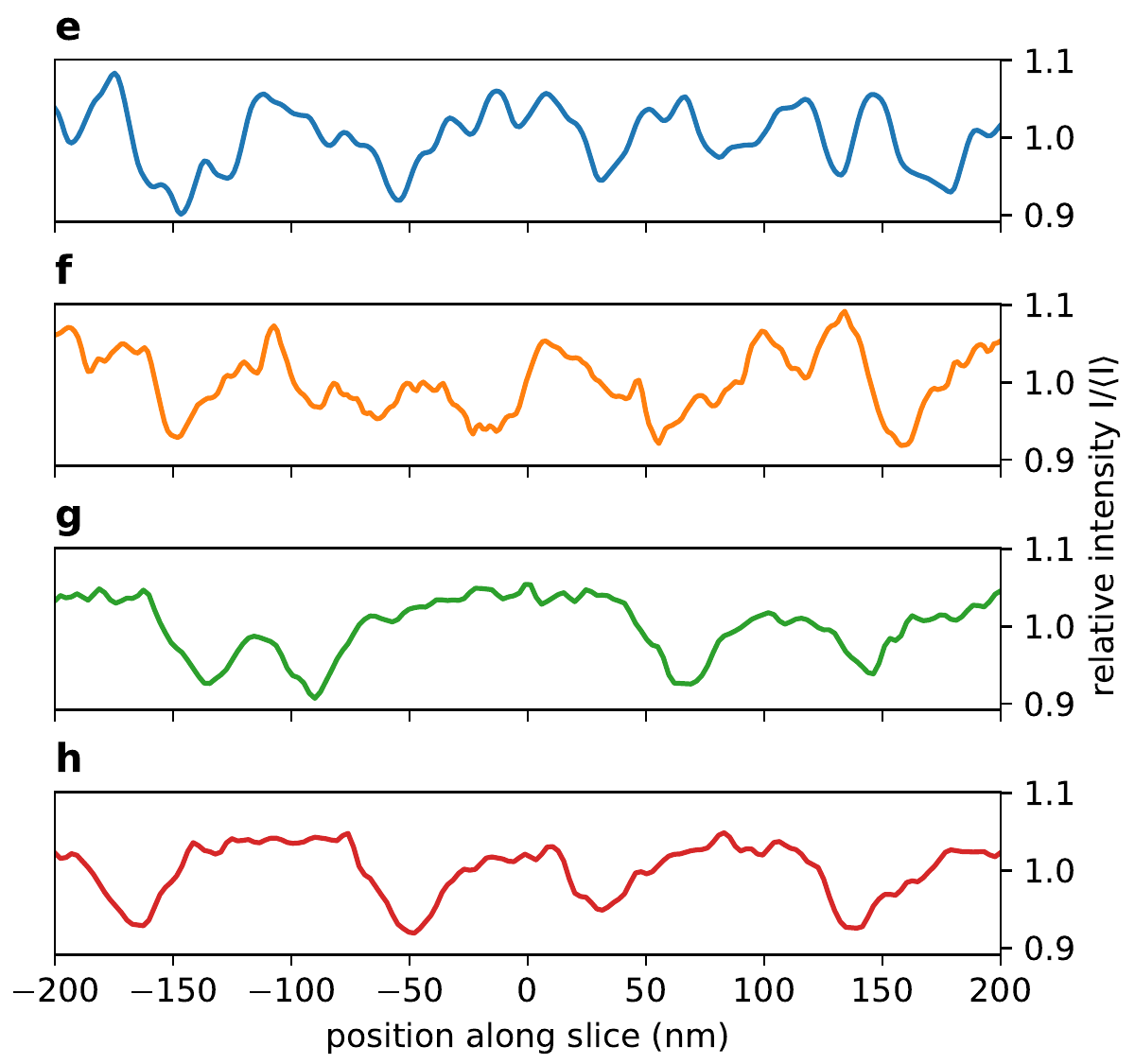}
\caption{\subf{a-d} Locations of the slices through stacking domain boundaries in several spectroscopic datasets. Scalebar applies to all panels and $E_0=38$\,eV for all images and all images are individually optimized for contrast. The epitaxial graphene datasets have an original resolution of 2.2\,nm/pixel, the TBG dataset a resolution of 3.7\,nm/pixel.
\subf{e-h} Normalized intensity along the slices indicated in respectively \subref{a-d}.
}
\label{fig:multislicelocs}
\end{figure}

Nevertheless, we compare the contrast as a function of $E_0$ as observed in the epitaxial graphene samples to the twisted case by appropriate cross-sections through domain boundaries. The cross-sections, shown in \figref{multislicelocs}, were taken through multiple domain boundaries, but without attempting to cross an AA-site, as any remaining drift would shift the cross-section away from the AA-site, invalidating such results.

\begin{figure*}[!ht]
\includegraphics[width=\textwidth]{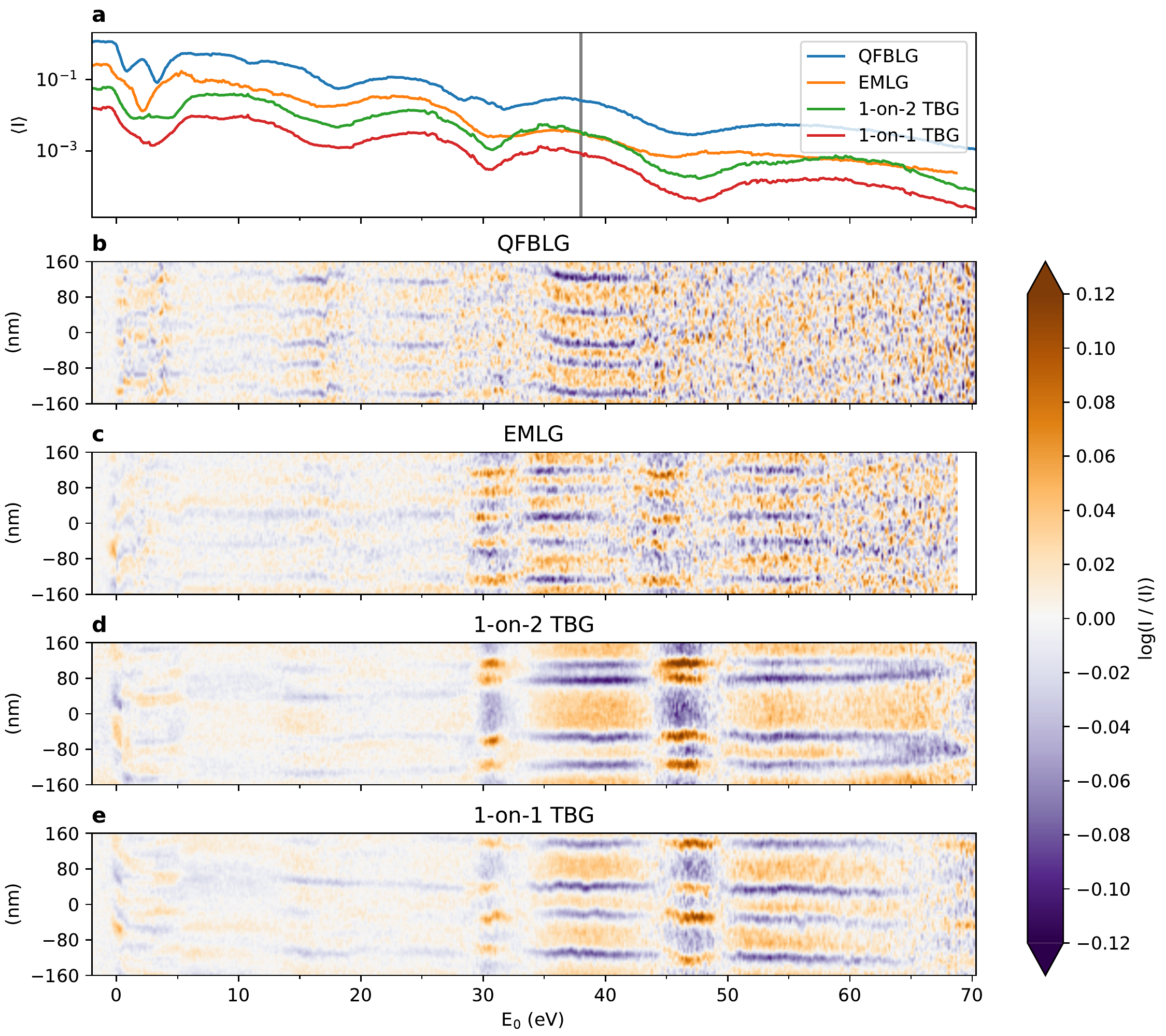}
\caption{\textbf{Comparison of domain boundaries with a single graphene top layer}: \subf{a} Average intensity along each slice $\langle I \rangle$ as a function of $E_0$, offset for clarity.
\subf{b-e} Log-contrast, i.e. (natural) logarithm of the intensity relative to the slice average $\langle I \rangle$ as a function of $E_0$. The SiC slices (QFBLG and EMLG) are taken in the same area of Ref.~\cite{de_jong_data_2018}, the TBG slices are taken from the data in Ref.~\cite{de_jong_data_2021}. Locations of the different slices are shown in \figref{multislicelocs}.}
\label{fig:multislice}
\end{figure*}

The resulting energy-dependent average reflectivity $\langle I\rangle(E_0)$ along each slice is shown in \figref[a]{multislice}, recovering the expected spectra for QFBLG, EMLG, bilayer graphene on hBN and trilayer graphene on hBN. 
The log-contrast $\log(I / \langle I\rangle)$ as a function $E_0$ along each slice is shown \figref[b-e]{multislice}. Here, in addition to the regular flat field correction (as described in Ref.~\cite{de_jong_quantitative_2020}), a linear profile along the spatial direction is subtracted to compensate for remaining illumination inhomogeneity.

Contrast is remarkably similar for all systems shown, with dark (blue) domain boundaries for $E_0$ between 35 and 43\,eV and contrast inversion above and below that, consistent with the calculations, which show similar contrast inversions.
For QFBLG and EMLG, the contrast washes out at higher $E_0$ (QFBLG above 45\,eV, EMLG above 65\,eV). However, this is an artefact most likely caused by insufficient integration time combined with incorrect focus tracking of the objective lens, causing the images to defocus at the high energies.
Notably, the contrast below 30\,eV is lower in EMLG than in the others, possibly due to the slightly different structure of the buffer layer compared to `true' lowest graphene layers in QFBLG and the TBG areas.

Some residual drift is present in the slices of each system, as the domain boundaries move collectively as a function of energy. Notably, some domain boundaries also move with respect to each other, e.g. the center two domain boundaries of 1-on-1 TBG around 39\,eV. Such dynamics of the moir\'e pattern are in fact common and have been characterized more precisely for TBG~\cite{de_jong_imaging_2021}.
By comparing the 1-on-1 TBG in \figref{multislice} to the unit cell averaged data in \figref{TBGcuts}, it becomes clear that the log-contrast for unit cell averaged data is about 1.5 times larger (0.2 peak-to-peak in \figref{multislice} versus 0.3 Bernal-to-peak in the unit cell averaged case).\footnote{In terms of non-log contrast this corresponds to approximately a factor 1.2 peak-to-peak for the slices and 1.35 Bernal-to-peak for the unit cell averaged case.}
Contrary to theory, all systems seem to consistently show at least some contrast for all energies lower than 30eV, although with varying strength and sign.

Domain boundaries in all four datasets are wider than the 6--11\,nm predicted by simulations~\cite{alden2013strain,lebedeva_two_2020}, even when taking into account the non-perpendicular cuts.
This suggests the data is again limited by electron optical reasons: either electron optical resolution of the measurements, or contribution of a phase component in addition to the pure amplitude component of the calculated stacking contrast to the image formation.

The 1-on-2 TBG data is remarkably similar to that of the other systems in this section, matching well to theory. The most evident difference in this system is the contrast between neighboring domains, which correspond to ABA and ABC stacking respectively, for example around 0, 10, 33 and\,65 eV. This contrast between different Bernal and rhombohedral stackings will be explored in more detail in the next section.

\begin{figure}[htb]
\includegraphics[width=\columnwidth]{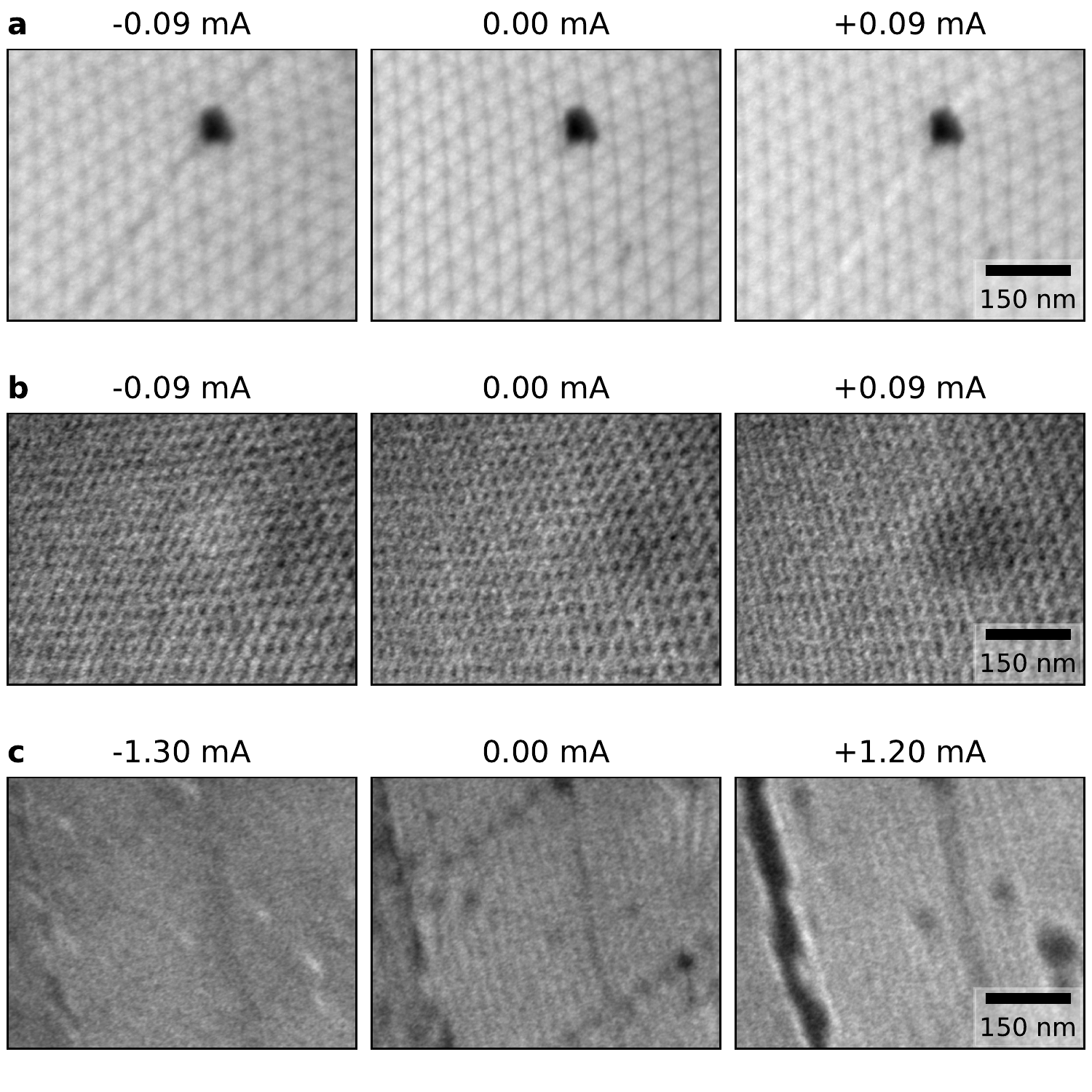}
\caption{\textbf{Defocus series.} 
\subf{a} Defocus series of TBG at $\theta\approx0.18^\circ$ at $E_0=36.5\,\text{eV}$. The contrast of a diagonal line feature, presumably in hBN substrate inverts: from dark in underfocus to bright in overfocus.
\subf{b} Defocus series of TBG at $\theta\approx0.6^\circ$ at $E_0=37.3\,\text{eV}$. A round feature, presumably a bubble under the TBG, inverts contrast from bright in underfocus to dark in overfocus.
\subf{c} Defocus series of graphene on SiC at $E_0=37.3\,\text{eV}$. Several adsorbed carbohydrate residue particles change from bright in underfocus to dark in overfocus.
Defocus is indicated above each panel in terms of objective lens excitation current relative to focus. Scalebars apply to all panels.
}
\label{fig:defocus_series}
\end{figure}

Further evidence that the contrast in 1-on-1 TBG and graphene on SiC is pure amplitude contrast is given by the defocus series shown in \figref{defocus_series}.
If there would be a (strong) phase component to the contrast, this would invert as a function of defocus. 
Indeed, for all three defocus series, there are features present of which the contrast does invert as a function of defocus, confirming these focus series cross the in-focus condition. However, the domain boundaries do not show any signs of inverting contrast as a function of defocus in any of them. This confirms a pure amplitude contrast for domain boundaries both in TBG and in graphene on SiC.

\section{Beyond bilayers}\label{sec:Gbeyondbilayers}

While for bilayer graphene as explored in the previous sections, both possible Bernal stackings (AB / AC) are strictly equivalent as they are related by rotational symmetry (ignoring substrate effects), for trilayer and more layers, this equivalence is broken. In this section, the consequences of this for BF LEEM imaging of stacking domains multilayer (i.e. more than two layers) graphene are explored.

Bernal stacked trilayer graphene (ABA, occurring in nature's graphite) has a distinct structure from rhombohedral graphene (ABC).
The latter is hypothesized to possess interesting electronic properties, including flat bands~\cite{pierucci_atomic_2016, henck_flat_2018,marchenko_extremely_2018} and a slightly different stacking energy~\cite{halbertal_moire_2021,guerrero-aviles_relative_2021}. 
However, large areas of rhombohedral graphene turn out to be hard to create using standard stacking methods and even harder to stabilize, with samples typically showing a strong tendency to revert to Bernal stacking~\cite{Yoo2019,guerrero-aviles_relative_2021}.

\begin{figure}[!ht]
\includegraphics[width=\columnwidth]{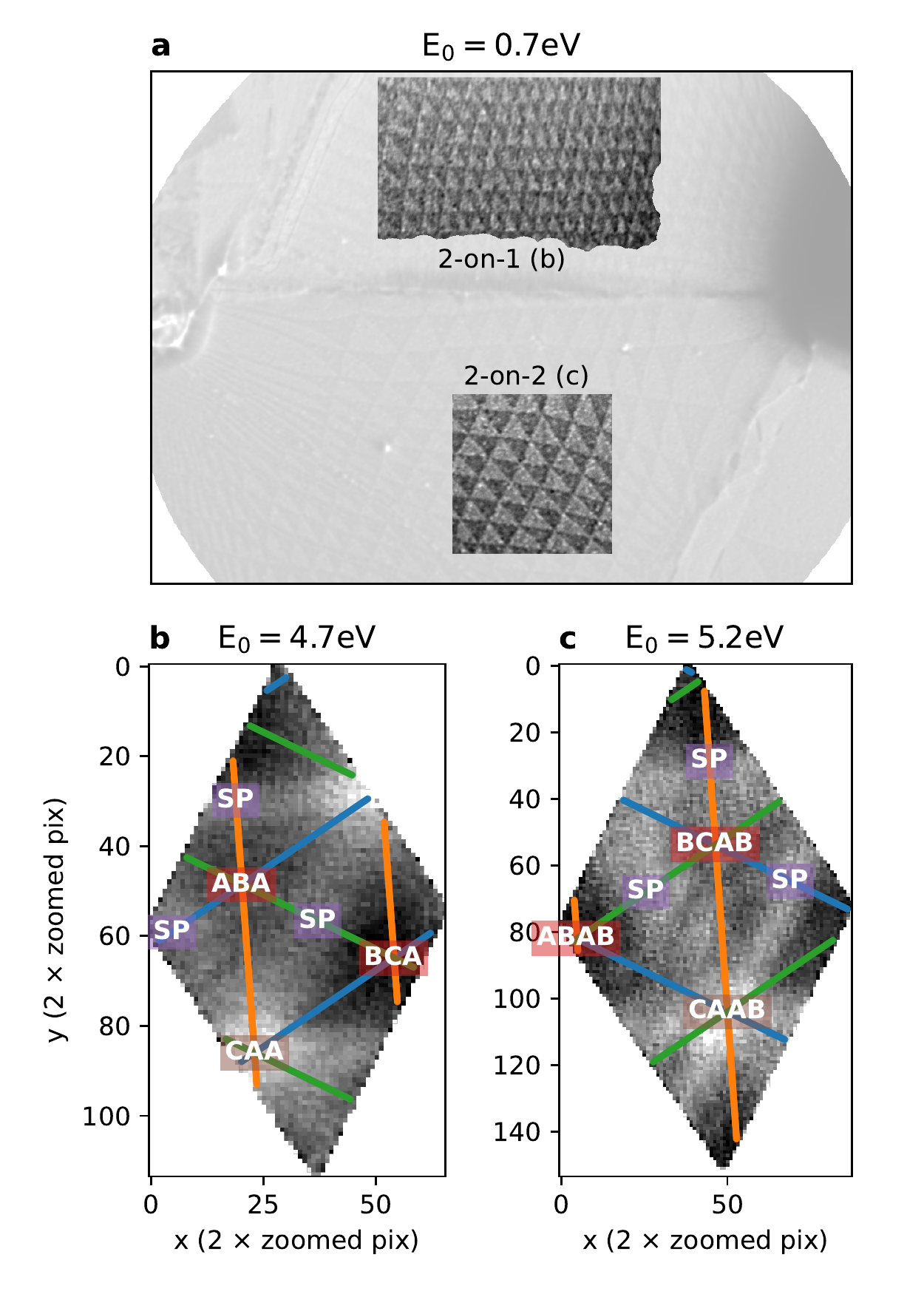}
\caption{\subf{a} BF LEEM image of an area of a TBG sample with both a 2-on-1 and a 2-on-2 area. The areas used for unit cell averaging are highlighted.
\subf{b} Average unit cell for the 2-on-1 area at $E_0=4.7$eV with the deduced stacking assignment indicated.
\subf{c} Average unit cell for the 2-on-2 area at $E_0=5.2$eV with the deduced stacking assignment indicated.
}\label{fig:multilayerunitcells}
\end{figure}

Both minimally twisted multilayers and strained epitaxial graphene form a natural platform to study differences between different stackings, as areas of different stackings are inherently created in alternating patterns. Furthermore, they are topologically protected, since boundary nodes, which are as such sometimes referred to as `twistons' in the twisted case, can only disappear by moving al the way to the edge of the sample.
This latter behavior corresponds to full untwisting of the sample over relatively large length scales for twisted samples. For the strainons in the strained epitaxial samples, the same holds, as the conservation is enforced by the binding to the substrate step edges and defects. 

\begin{figure*}[!ht]
\includegraphics[width=\textwidth]{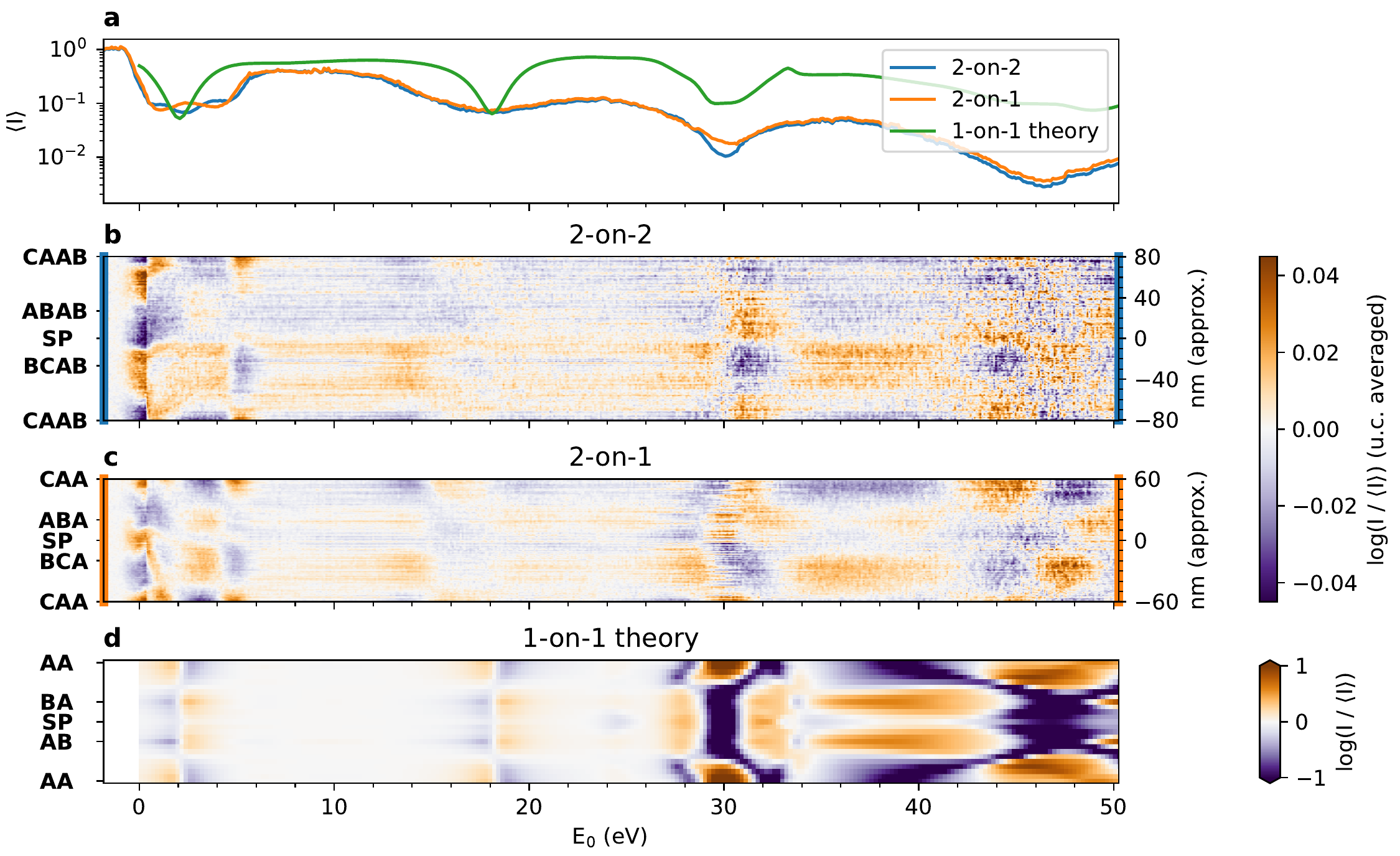}
\caption{\subf{a} Average intensity $\langle I \rangle$ as a function of $E_0$ for the areas indicated in \figref[a]{multilayerunitcells} and the average intensity for the computed theoretical reflectivities. \subf{b,c} Relative intensity of cuts through the averaged unit cells in the three directions indicated in \figref[b,c]{multilayerunitcells} for twisted bilayer-on-bilayer graphene. Data was taken at a magnification of 3.7\,nm/pixel. Note the much smaller range of the colormap compared to \figref{multislice} and \ref{fig:TDBGcutsphase} (in the next section).
\subf{d} Calculated relative intensity scaled by the average intensity for 1-on-1 bilayer.}\label{fig:multilayeraverages}
\end{figure*}

Aside from DF-LEEM, which can be used  to distinguish the different possible stackings in trilayer graphene on SiC~\cite{dejong2018intrinsic}, we here explore the BF-LEEM characteristics of both domains and domain boundaries of different trilayer and quadlayer stackings.

As visible in \figref[d]{multislice} (in the previous section), for 1-on-2, the domain boundaries yield very similar contrast to 1-on-1. This is expected, as the `substrate' (an extra layer of graphene on hBN versus hBN in this case) has much less influence on the observed LEEM spectra than the top layers. 
However, some contrast between ABA and ABC stacking does appear when comparing to the bilayers, in particular around $E_0\approx 10$\,eV and $E_0\approx 65$\,eV, confirming the broken rotational symmetry. 

Wildly different is the bright field contrast for samples where the twisted top layer consists of bilayer graphene, i.e. 2-on-1 and 2-on-2 TBG, as shown for $E_0=0.7$\,eV in \figref[a]{multilayerunitcells}. Here, it is already clear that Bernal versus rhombohedral stacking dominates the contrast near mirror mode, visible as dark and bright triangles. These triangles are used to compute $\vec u (\vec r)$ for unit cell averaging.

When looking at the resulting energy-dependent, unit cell averaged 2-on-1 and 2-on-2 data shown in \figref{multilayeraverages}, the difference in contrast compared to the 1-on-X data in the previous section is clear. The overall contrast is much lower and the contrast between ABA and ABC stacking dominates, although some (C)AA(B) and SP contrast is visible, for example around 5\,eV.

\subsection{2-on-2 graphene layers: phase contrast}

\begin{figure*}[!ht]
\includegraphics[width=0.76\textwidth]{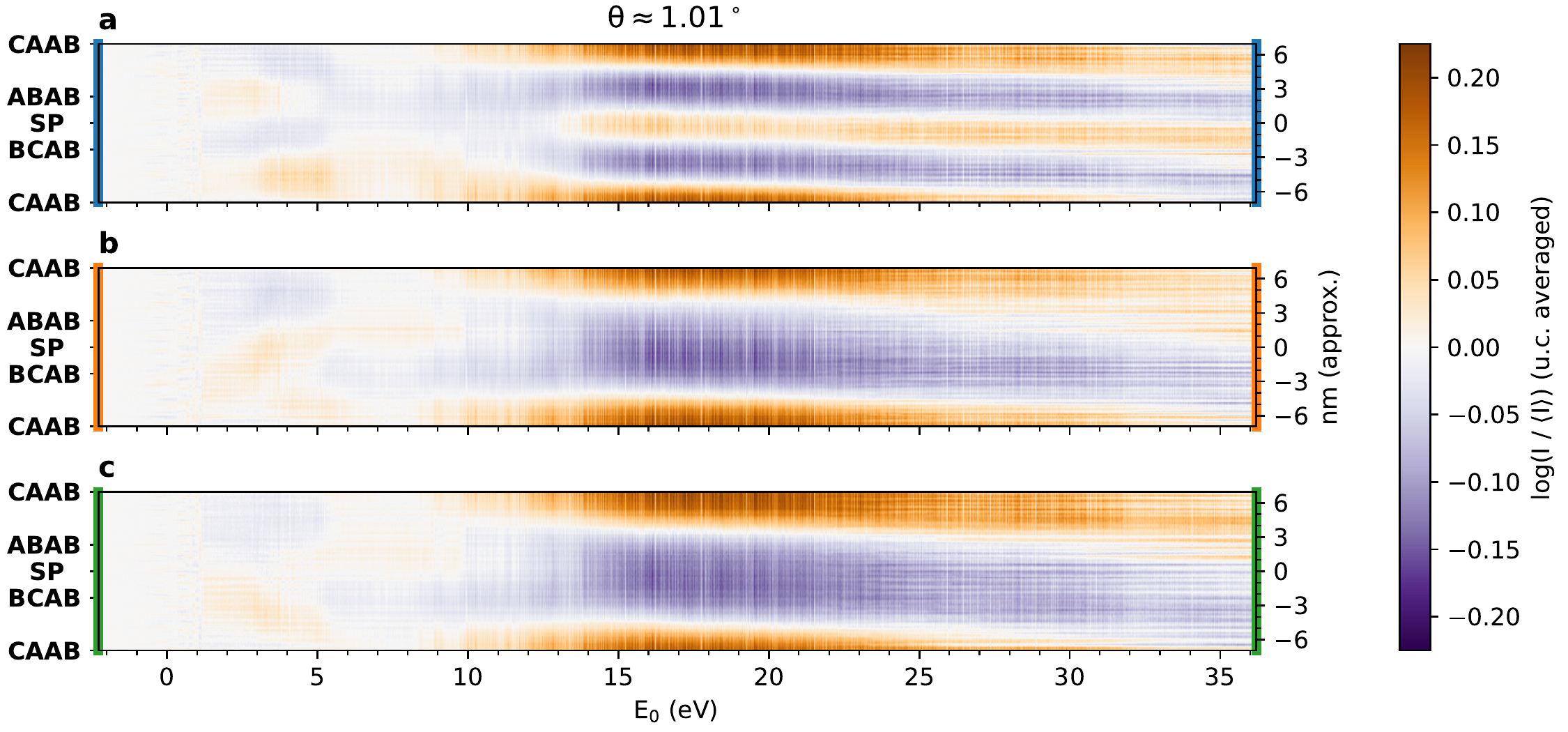}
\includegraphics[width=0.23\textwidth]{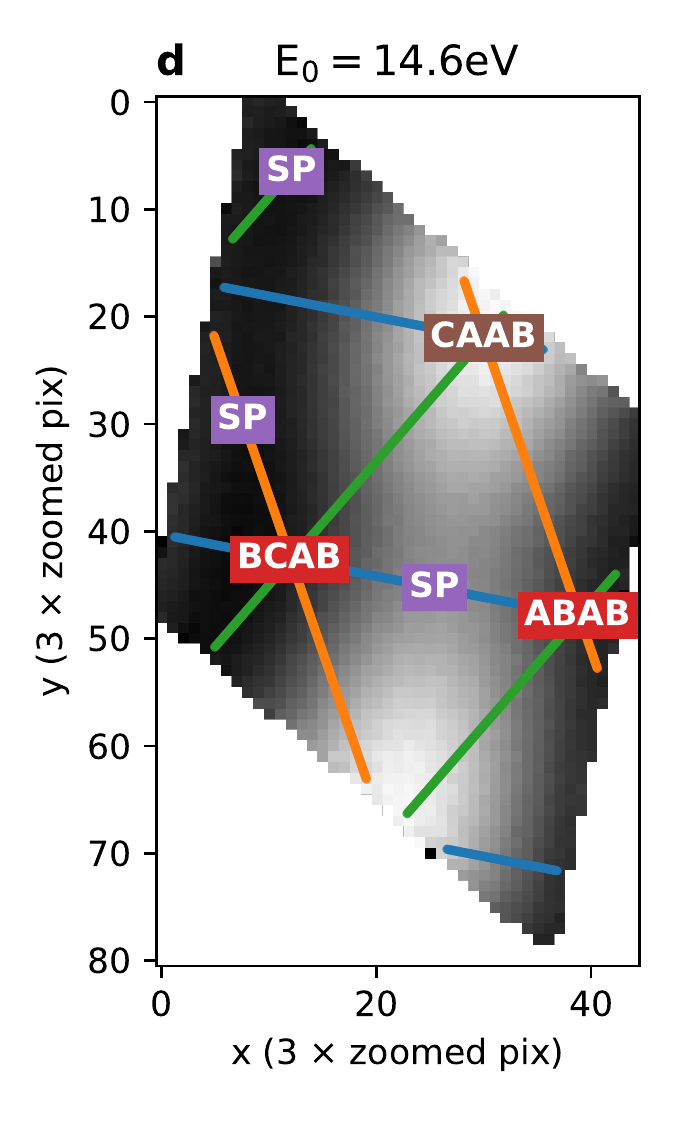}
\caption{\subf{a-c} Relative intensity of cuts through the averaged unit cells for twisted 2-on-2 layer graphene near the magic angle. Data was taken at a magnification of 0.9\,nm/pixel. Note the difference in colorscale compared to Figure \ref{fig:multilayeraverages}.
\subf{d} Averaged unit cell with the cuts taken in \subref{a-c} indicated. Assignment of CAAB is arbitrary, see main text.}\label{fig:TDBGcutsphase}
\end{figure*}

The results shown in the previous sections are fairly consistent with the calculations and therefore with pure amplitude contrast.
However, something unexpected happens for 2-on-2 TBG data of higher twist-angle, and thus smaller unit cell area, such as in \figref{TDBGcutsphase}.

Although the size of this moir\'e is close to the resolution limit of the instrument, the contrast is very high and shows no inversions between $\sim 10$\,eV and 36\,eV. 
The observed contrast is the highest of all measurements presented in this work, peaking at $\left.\frac{I_\text{max}}{I_\text{min}}\right|_{E_0} \approx 1.5$ for a relatively wide region around $E_0 = 20$\,eV.\footnote{$1.5 \approx \exp(0.42)$, i.e. the contrast of 1.5 corresponds to difference of 0.42 on the color scale in the figures.}

\begin{figure*}[!ht]
\includegraphics[width=0.95\textwidth]{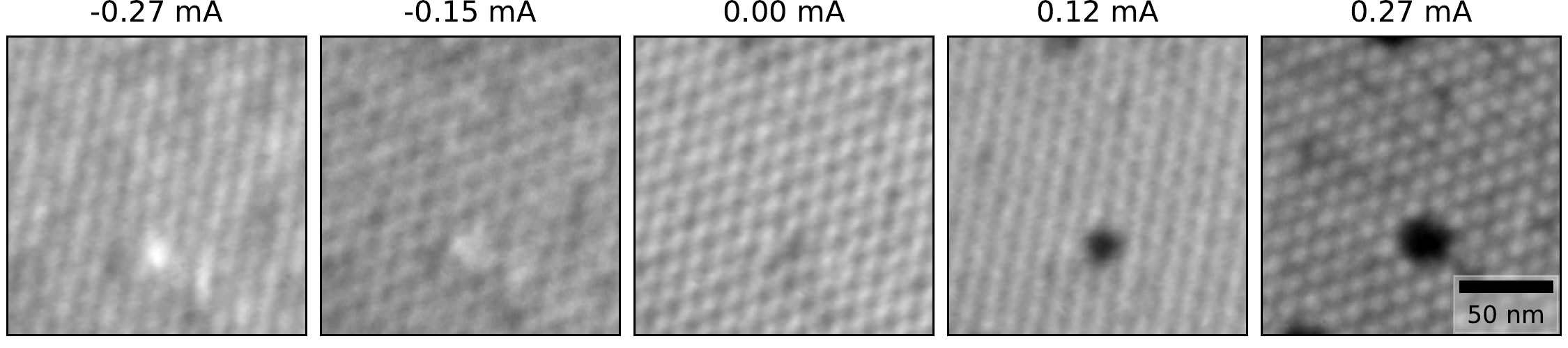}
\caption{\textbf{Defocus series of $\theta=1.01^\circ$ 2-on-2 TBG.} Data is taken at $E_0=5.3\,\text{eV}$. Defocus relative to the center panel is expressed in mA excitation of the objective lens.}\label{fig:defocus_large_angle}
\end{figure*}

In this limit of small domains, with three inequivalent sublattices (BCAB, ABAB and CAAB have three different intensity contributions), no definitive stacking assignment can be made from the experimental data.
The stacking assignment as indicated in \figref[d]{TDBGcutsphase} is therefore just an indication: the brightest point need not correspond to CAAB in this case and the slices might be shifted (but not rotated) relative to the actual bernal and AA-node points.
Nevertheless, the much higher contrast and lack of contrast inversion at this higher twist angle compared to the $\theta=0.08$ data (In \figref[b]{multilayeraverages}, note the difference in color scale maximum.), indicates phase contrast in addition to amplitude contrast (where electrons reflecting off different parts of the unit cell interfere with each other) dominates for these higher twist angles in 2-on-2 TBG.

To further corroborate the phase contrast, a defocus series is shown in \figref{defocus_large_angle}.
The same small size of the moir\'e pattern that would enable phase contrast puts it however right on the edge of the achievable resolution in the LEEM.
It is convoluted with some remaining astigmatism and sample drift, but the observed defocus series shows contrast shifting from dark dots in a triangular grid on a bright hexagonal background to bright dots on a darker hexagonal background.
This shift of contrast as a function of defocus combined with the fact that the contrast is virtually independent of $E_0$ for $E_0 > 10$\,eV, leads us to conclude that the observed contrast is indeed due to phase contrast.

Although it is convoluted with some remaining astigmatism and sample drift, the observed defocus series combined with the fact that the contrast is virtually independent of $E_0$ for $E_0 > 10$\,eV, leads us to conclude that the observed contrast is indeed due to phase contrast.

\section{Moiré metrology}\label{sec:moiremetrology}

\begin{figure*}[!ht]
\includegraphics[width=\textwidth]{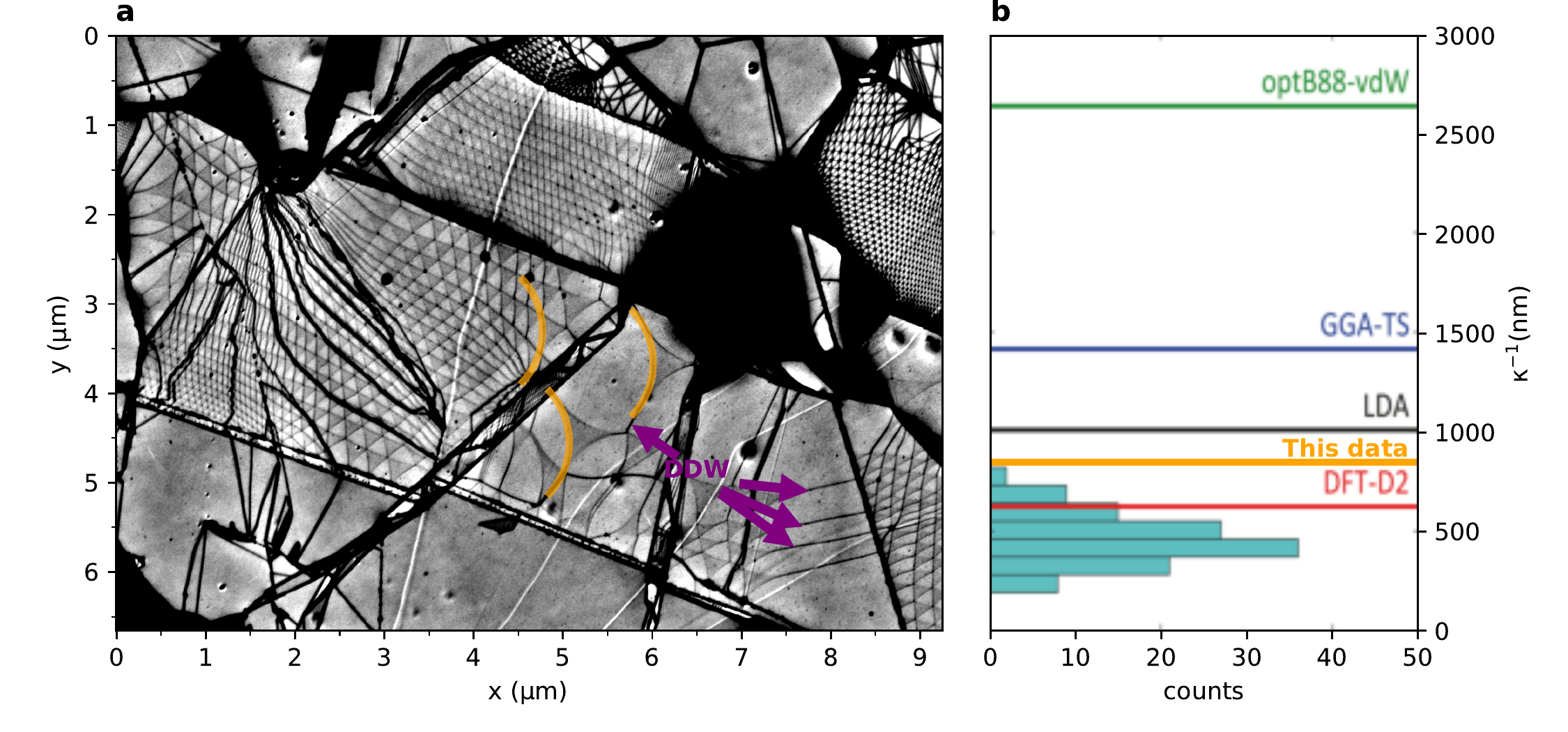}
\caption{\textbf{Moir\'e metrology:} \subf{a} BF-LEEM image of a 2-on-2 TBG area with very low twist angle  (same device as ref.~\cite{de_jong_imaging_2021}). Domain boundaries in the 2-on-2 TBG areas are clearly visible, as is a slight contrast between ABAB and BCAB domains. In the very low twist angle areas, the curvature of the domain bounderies is very apparent. For three of them a matching orange arc with a curvature of $ \kappa^{-1} = 850$\,nm, is overlayed. Some double domain walls (DDW) are indicated with purple arrows. \subf{b} Measured curvatures using SNOM as a histogram with the predictions from different ab-initio calculation schemes indicated as lines, adapted from~\cite{halbertal_moire_2021}. The curvature drawn in \subref{a} is also indicated with an orange line.}\label{fig:moiremetrology}
\end{figure*} 

Beyond measuring the contrast of reflected low energy electrons of moir\'e patterns and determining the local twist angle, there is more that we can learn from imaging moir\'e patterns in such samples.

As described by Halbertal et al. for the case of 2-on-2 layer twisted graphene~\cite{halbertal_moire_2021,enaldiev_stacking_2020}, the shape of the domain boundaries can be directly related to any energy differences between different stackings and therefore can be used to \textit{measure} (hence moir\'e \textit{metrology}) these stacking energy differences. 

In general, in a system with states of different energy that is in thermal equilibrium, the state with the lower energy will occur more often. The ratio between occupancy of the states is directly related to the energy difference by the Boltzmann factor.
Although the number of twistons in a twisted system, and therefore the number of alternating domains is conserved (ignoring edge cases), the size of the domains can change by movement of the domain boundaries. 

However, the relative size of different stacking domains does not map directly to such a Boltzmann factor, as the energy cost per unit length of domain boundary has to be taken into account. 
What is more, this energy cost is dependent on the local angle between the domain boundary and the atomic lattice. 
Nevertheless, Halbertal et al. show that the generalized stacking fault energy (GSFE), the stacking energy as a function of relative displacement of lattices, can be directly related to the curvature $\kappa$ of domain boundaries of the triangular domains, which they image using scanning near-field optical microscopy (SNOM). This methodology works for 2-on-2 TBG, but also for other materials.

As shown in the preceding sections, LEEM can similarly image domains in diverse systems of heterostacks, providing another way to measure the shapes of these domain boundaries and therefore calibrate theoretical calculations of such stacking differences.

As calculations suggest that both magnitude and direction of heterostrain influence the energy differences between different stackings, measuring larger areas of twisted heterostructures is an effective means to measure those effects~\cite{guerrero-aviles_relative_2021}.  
In such samples, varying strain can be characterized locally using GPA (as described in Ref.~\cite{mesple_heterostrain_2021,halbertal_moire_2021,de_jong_imaging_2021,halbertal_extracting_2022}) and in conjunction the energy difference between the stackings can be determined by domain boundary curvature. 
This way, varying strain and energy difference can be connected experimentally.

In \figref{moiremetrology} a proof-of-concept of using LEEM to do such measurements is shown. Although the sample used only showed some areas of low enough twist angle to measure $\kappa$, it is already clear that we measure a value outside of the range of values that Halbertal et al. obtained as indicated by the histogram in \figref[b]{moiremetrology}. Interestingly, the value we find is closer to theoretically predicted values using LDA, GGA-TS and optB88-vdW, but farther away from the one from DFT-D2 (for more details on the differences between these calculations, see the Methods section of Ref.~\cite{halbertal_moire_2021}).
Furthermore we observe double domain walls in the 2-on-2 TBG (for example the ones indicated with purple arrows in \figref{moiremetrology}), similar to observations by Halbertal et al., although we note that these did not occur in the 1-on-1 and 2-on-1 areas of the sample.

The possibilities for such measurements in a LEEM opens up a further research avenue: to explore the \textit{dynamics} of the domain wall positions in such minimally twisted samples, similar to the work on higher twist angle data in Ref.~\cite{de_jong_imaging_2021}. By mapping the domain wall mobility as well as equilibrium curvatures as a function of temperature, it would be possible to not only explore the energy differences between the stackings, but also further characterize the stacking energy landscape.

\section{Conclusion}

In conclusion, we have shown that for large stacking domains in bilayer graphene, the local stacking in the domain walls and nodes is the primary BF-LEEM amplitude contrast mechanism for $E_0 \gtrsim 30$\,eV.
The contrasts observed in this energy range correspond very well to theoretical calculations, both for (low angle) 1-on-1 and 1-on-2 twisted bilayer graphene as well as for QFBLG and EMLG on silicon carbide, although the observed contrast is much lower due to the spatial resolution limitations of the experiment and thermal broadening.

Furthermore, we have applied similar methods to map the stacking contrast for 2-on-2 and 2-on-1 TBG. Here, for low angle data, the contrast is much lower, and mostly caused by contrast between the (meta-)stable Bernal and rhombohedral stackings, with domain boundaries only exhibiting minor contrast at some landing energies.
Curiously, for $\theta \approx 1^\circ$, 2-on-2 TBG exhibits a much stronger contrast, stronger even than 1-on-1 TBG, suggesting that a phase contrast mechanism distinct from the local stacking contrast starts to become dominant.

Nevertheless, it is also clear that there are still algorithmic limitations of the current implementation of the unit cell averaging, both in the unit cell averaging itself and in the adaptive Geometric Phase Analysis (GPA) used to obtain the displacement field $\vec u(\vec r)$. 
As noted before, edges of the unit cell can be treated more accurately. 
Furthermore, some residual drift is clear from the asymmetry of the different cuts, e.g. in \figref{TBGcuts}. Although this is not correctable by regular drift correction, the three slices could be symmetrized to correct for this. Finally, the fluctuations of the moir\'e pattern can be taken into account by computing $\vec u (\vec r)$ for several different landing energies and interpolating between those for the unit cell averaging.

However, from the results obtained here we can draw several conclusions.
The optimal landing energy range to image domain boundaries in a bilayer of graphene, seems to be 30-- 50\,eV, where a strong amplitude contrast occurs and the intensity is still relatively high.
For domain boundaries between deeper lying layers, the amplitude contrast at high values of $E_0$ is much lower, and the optimal energy to image the domains themselves is at very low energies, 0 -- 10\,eV, where there is plenty of intensity and the work function difference causes relatively strong contrast.
An exception holds for larger twist angles / smaller domains, where phase contrast is the dominant contrast mechanism causing strong contrast between 10 and 20\,eV.
We speculate that these trends are more generally applicable to stacking boundaries in Van der Waals heterostacks, beyond the graphene-graphene system alone: 
(i) Significant amplitude stacking contrast only for $E_0$ larger than the energy at which the first order diffraction spots appear, (ii) large deeper lying domains most clearly imagable by slight work function differences and (iii) small domains dominated by phase contrast, especially for deeper lying stacking differences.

The contrast mechanisms as investigated here are exploited to measure local strain and twist angle in TBG in Ref.~\cite{de_jong_imaging_2021} and to explore relative strain and disorder in epitaxial graphene on SiC in Ref.~\cite{de_jong_2022_morphology}.

Finally, we have shown the potential of using such contrast in twisted heterostacks to closely study the energy differences between different possible stackings.

\section*{Acknowledgements}
We thank Marcel Hesselberth and Douwe Scholma for their indispensable technical support.
We thank Christian Ott and Heiko Weber for the fabrication of the graphene on SiC samples. This work was supported by the Netherlands Organisation for Scientific Research (NWO/OCW) as part of the Frontiers of Nanoscience program. It was also supported by the Spanish Ministry of Science, Innovation and Universities (Project No. PID2019-105488GB-I00).
\bibliography{domainboundaries.bib}
\end{document}